\documentclass{PoS}

\title{Taste symmetry and QCD thermodynamics\\ with improved staggered fermions}

\ShortTitle{Taste symmetry and QCD thermodynamics
}

\author{\speaker{Alexei Bazavov}$^a$\footnote{Present address:
Physics Department, Brookhaven National Laboratory, Upton, NY 11973, 
USA}
{ } and Peter Petreczky$^b$
\footnote{This work has been supported in part by 
          contracts DE-AC02-98CH10886
          and DE-FC02-06ER-41439
          with the U.S. Department of Energy
          and contract 0555397 with the National Science
          Foundation. The numerical calculations have been performed
          using the USQCD resources at Fermilab and JLab, the BlueGene/L
          at the New York Center for Computational Sciences (NYCCS)
          and the BlueGene/L at the J\"{u}lich Supercomputing Center.
          We thank Z.~Fodor and S.~Katz for sending us the stout data.}
         { }[for the HotQCD collaboration]
\footnote{HotQCD Collaboration members are:
A.~Bazavov,
T.~Bhattacharya,
M.~Cheng,
N.H.~Christ,
C.~DeTar,
S.~Gottlieb,
R.~Gupta,
P.~Hegde,
U.M.~Heller,
C.~Jung,
F.~Karsch,
E.~Laermann,
L.~Levkova,
C.~Miao,
R.D.~Mawhinney,
S.~Mukherjee,
P.~Petreczky,
D.~Renfrew,
C.~Schmidt,
R.A.~Soltz,
W.~Soeldner,
R.~Sugar,
D.~Toussaint,
W.~Unger,
P.~Vranas}\\
\llap{$^a$}Department of Physics, University of Arizona,
        Tucson, AZ 85721, USA\\
\llap{$^b$}Physics Department, Brookhaven National Laboratory, Upton, NY 11973, 
USA\\
}

\abstract{
Taste symmetry violations in staggered fermion formulations correlate strongly with the cut-off (lattice spacing) dependence in thermodynamic quantities. Better taste symmetry on the lattice can be achieved either by decreasing the lattice spacing and going to larger temporal extent in finite-temperature calculations, or by further improving the action. The highly improved staggered quark (HISQ) action offers a further degree of improvement and substantially reduces taste violations. We report on our studies of the 2+1 flavor QCD thermodynamics with the HISQ/tree action. By systematically comparing HISQ/tree, asqtad, p4 and stout calculations we quantify how the cut-off effects manifest themselves in different thermodynamic quantities, including the renormalized Polyakov loop, chiral condensate, various fluctuations and correlations of conserved charges. The implications for the equation of state and a comparison to the hadron resonance gas model are also discussed. The chiral aspects
of the finite-temperature transition are discussed in the companion HotQCD
contribution~\cite{WSlat10}.
}

\FullConference{The XXVIII International Symposium on Lattice Filed Theory\\
		 June 14-19,2010\\
		 Villasimius, Sardinia, Italy}

\begin{document}

\section{Introduction}

Improved staggered fermion formulations are widely used to study QCD
at non-zero temperature and density \cite{carleton}. Staggered
fermions preserve a part of the chiral symmetry of continuum QCD
and are inexpensive to simulate numerically. After staggering the fermion
field the number of doublers is reduced from sixteen  (in four
dimensions) to four. These four species of fermions, that we call ``tastes'',
are inequivalent at non-zero lattice spacing, because the taste
symmetry is broken. It is restored in the continuum limit.
To reduce the number of fermions further the rooting procedure is
used. Its validity is discussed in Ref.~\cite{sharpe06creutz07}.

Violations of the taste symmetry at non-zero lattice spacing contribute
to the cut-off dependence of different physical observables. To reduce
the taste violations, smeared links, \textit{i.e.} weighted averages of 
different paths on the lattice that connect neighboring points, 
are used in the staggered Dirac operator. Various improved staggered
fermion formulations, like p4, asqtad, stout and HISQ differ in the choice
of smeared gauge links. The highly improved staggered quark (HISQ)
action was developed by the HPQCD/UKQCD collaboration~\cite{Follana:2006rc}
and has been shown to offer a very high degree of improvement in the
taste symmetry~\cite{Follana:2006rc,Bazavov:2010ru}.
This action together with the one-loop tadpole improved gauge action
is used by the MILC collaboration~\cite{Bazavov:2010ru} to generate 
a library of ensembles
with several lattice spacings and light quark masses. Such a combination
of the fermion and gauge actions was originally referred to as HISQ.
In our thermodynamic studies we use a tree-level improved gauge action.
Thus, to distinguish our variant of the action from the one that
MILC and HPQCD originally used we call it the HISQ/tree action.

In this contribution we extend 
our studies~\cite{hisqthermo}
of 2+1 flavor QCD thermodynamics with the HISQ/tree action,
compare with the results obtained with p4, asqtad and stout and
discuss the discretization effects in different thermodynamic quantities.

\section{Taste symmetry of different actions}\label{sec_tastes}

We have generated a library of zero- and finite-temperature HISQ/tree
ensembles along two lines of constant physics (LCP): $m_l=m_s/5$
and $m_l=m_s/20$, where $m_l$ ($m_s$) is the light (strange) quark
mass. The parameters of these ensembles have been reported
in~\cite{hisqthermo}.
For $m_s/5$ our lattice spacings range from
$a=0.15$ fm to 0.23 fm, and for $m_s/20$ from 0.08 fm to 0.23 fm.
The statistics has been updated and the current status is shown in
Table~\ref{tab_runs}.
\begin{table}
\centering
\small
\begin{tabular}{|l|l|l||l|r||r|r|}
\hline
\multicolumn{3}{|c||}{
 
} &
\multicolumn{2}{c||}{
$T=0$
} &
\multicolumn{2}{c|}{
$T>0$
} \\\hline
$\beta$ & $r_0/a$   & $am_s$ &$N_s^3\times N_\tau$& TU   & TU, $N_\tau=8$ & TU, $N_\tau=6$ \\\hline
6.000   & 2.094(21) & 0.1138 & $24^3\times32$ & 2,305 &       &  6,020 \\
6.050   & 2.198*    & 0.1064 & $24^3\times32$ & 2,675 &       &  6,340 \\
6.075   & 2.253*    & 0.1036 &                &       &       &  4,700 \\
6.100   & 2.310*    & 0.0998 & $28^3\times32$ & 1,875 &       &  6,280 \\
6.125   & 2.368*    & 0.0966 &                &       &       &  9,990 \\
6.150   & 2.427*    & 0.0936 &                &       &       & 11,220 \\
6.175   & 2.489*    & 0.0906 &                &       &       & 10,860 \\
6.195   & 2.531(24) & 0.0880 & $32^4$         & 2,365 & 9,990 & 11,100 \\
6.215   & 2.589*    & 0.0862 &                &       &       &  6,390 \\
6.245   & 2.668*    & 0.0830 &                &       & 1,600 &  6,410 \\
6.260   & 2.707*    & 0.0810 &                &       & 1,790 &        \\
6.285   & 2.775*    & 0.0790 & $32^4$         & 2,300 & 6,190 &  6,750 \\
6.315   & 2.858*    & 0.0760 &                &       & 1,805 &        \\
6.341   & 2.932*    & 0.0740 & $32^4$         & 1,325 & 7,020 &  6,590 \\
6.354   & 2.986(41) & 0.0728 & $32^4$         & 2,295 & 5,990 &  5,990 \\
6.390   & 3.076*    & 0.0694 &                &       & 8,760 &        \\
6.423   & 3.189(22) & 0.0670 & $32^4$         & 2,295 & 5,990 &  5,990 \\
6.460   & 3.292*    & 0.0640 & $32^3\times64$ & 1,030 & 10,990&        \\
6.488   & 3.395(31) & 0.0620 & $32^4$         & 2,295 & 11,990&  8,790 \\
6.515   & 3.470*    & 0.0604 & $32^4$         & 2,045 & 10,100& 10,430 \\
6.550   & 3.585(14) & 0.0582 & $32^4$         & 2,295 & 11,990&  7,270 \\
6.575   & 3.674*    & 0.0564 & $32^4$         & 2,295 & 14,500&  7,330 \\
6.608   & 3.765(23) & 0.0542 & $32^4$         & 2,295 & 11,990&  6,560 \\
6.664   & 3.994(14) & 0.0514 & $32^4$         & 2,295 & 11,990&  8,230 \\
6.800   & 4.568(30) & 0.0448 & $32^4$         & 2,295 & 11,990&  7,000 \\
6.950   & 5.187(39) & 0.0386 & $32^4$         & 2,295 & 11,990&  7,480 \\
7.150   & 6.186(72) & 0.0320 & $32^4$         & 2,295 & 11,990&  4,770 \\
\hline
\end{tabular}
\caption{The parameters of the numerical simulations:
gauge coupling, strange quark mass and the number of time
units (TU) for each run. The lattices are separated by 5 TU at
zero and 10 TU at finite temperature.
The volumes for zero-temperature runs are given in column 4.
At finite temperature we used $32^3\times8$ and $24^3\times6$ volumes.
In the second column the values with error bars represent $r_0/a$
determination from the static quark anti-quark potential and the values
marked by asterisk (*) are determined from an Allton-style fit.
The lattice spacing is determined by using $r_0=0.469$ fm.
}
\label{tab_runs}
\vspace{-5.5mm}
\end{table}

Violations of the taste symmetry lead to splitting of the staggered
pion spectrum into sixteen pion states grouped into eight multiplets,
that we label in the taste basis as
$\gamma_5$, $\gamma_0\gamma_5$, $\gamma_i\gamma_5$, 
$\gamma_i\gamma_j$, $\gamma_i\gamma_0$, $\gamma_i$, $\gamma_0$ and
$1$. $\gamma_5$ corresponds to the Goldstone pion, the lowest
state whose mass approaches zero
in the chiral limit, while the masses of the other tastes always
stay finite at non-zero lattice spacing. In the continuum the pion
spectrum becomes degenerate, and the difference in mass between tastes
vanishes. We denote the 
quadratic splittings as $\Delta_\alpha=m^2_\alpha-m_{\gamma_5}^2$,
where $\alpha$ is one of the above tastes except $\gamma_5$.

\begin{figure}
\begin{center}
\includegraphics[width=0.48\textwidth]{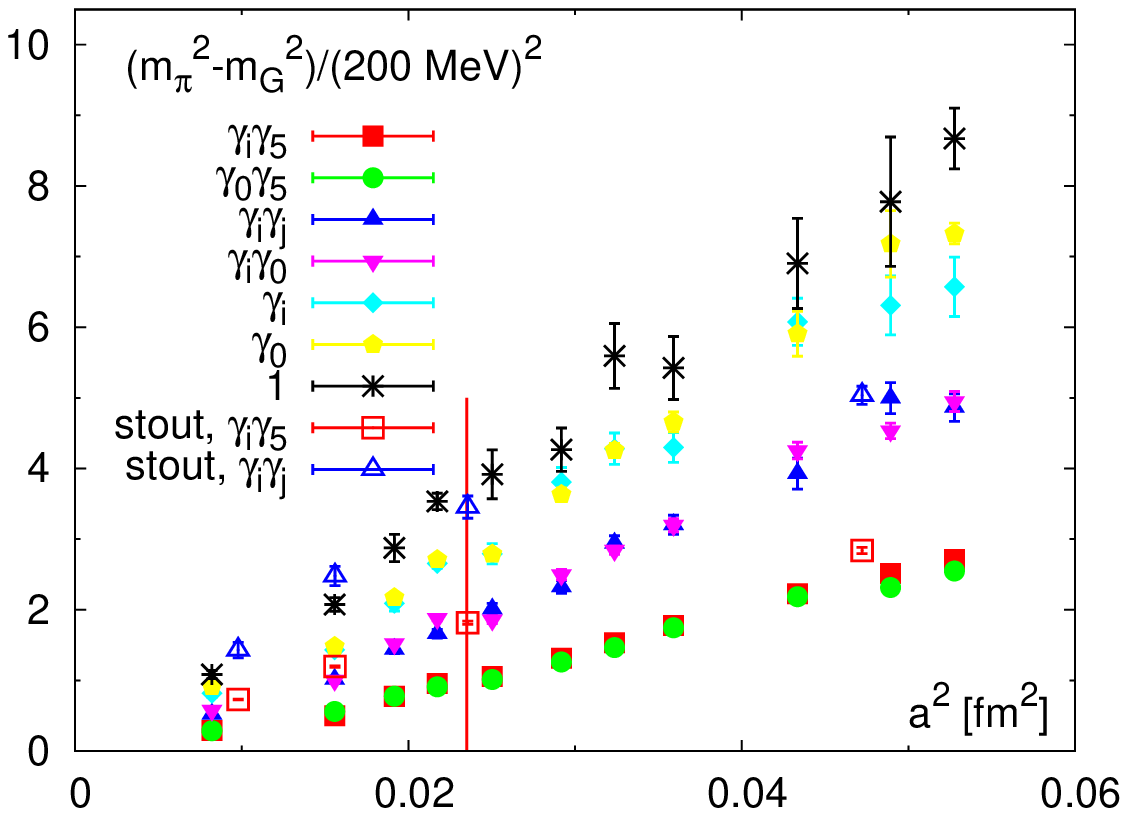}
\hfill
\includegraphics[width=0.48\textwidth]{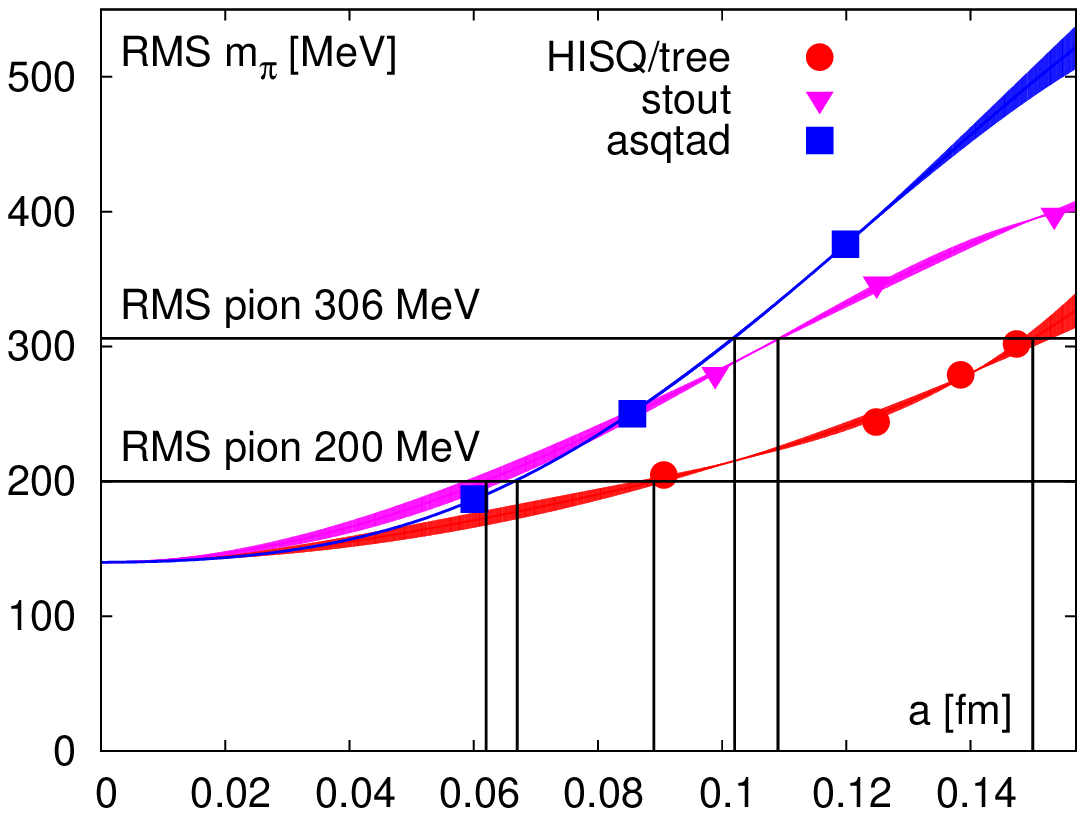}
\parbox[b]{0.48\textwidth}{
\caption{The splitting between different pion multiplets
calculated for HISQ/tree, $m_G\equiv m_{\gamma_5}$.}
\label{fig:pion_split}
}
\hfill
\parbox[b]{0.48\textwidth}{
\caption{RMS pion mass when $m_{\gamma_5}=140$ MeV. See details
in the text.}
\label{fig:pion_mass}
}
\end{center}
\vspace{-3mm}
\end{figure}

In earlier work~\cite{hisqthermo} we measured the pion
splittings on the ensembles with the light quark mass
of $m_s/5$. Zero-temperature ensembles along this LCP
were limited to lattices down
to $a=0.15$ fm. To study the behaviour of the spectrum at smaller
lattice spacings we measured $\Delta_\alpha$ using the valence light
quark mass of $m_s/5$ on zero-temperature ensembles with the light
sea quark mass of $m_s/20$. This is justified because quadratic
splittings do not depend on the quark mass at the lowest order in
the chiral perturbation theory, and also, as the measurements with
valence HISQ on asqtad sea have shown, $\Delta_\alpha$ are sensitive to the 
``roughness'' of configurations and not to the details of the action
used to generate them.
The results are presented in 
Fig.~\ref{fig:pion_split}. The vertical line separates our 
previously published~\cite{hisqthermo} data
(to the right) and the recently added calculations (to the left).
The stout data for $\Delta_{\gamma_i\gamma_5}$ 
and $\Delta_{\gamma_i\gamma_j}$ are shown as open symbols. 
We have also extracted approximate values of the stout 
$\Delta_{\gamma_i}$ and $\Delta_{1}$ from the right panel
of Fig.~2 in Ref.~\cite{Borsanyi:2010bp}, and assumed for stout
$\Delta_{\gamma_0\gamma_5}=\Delta_{\gamma_i\gamma_5}$,
$\Delta_{\gamma_i\gamma_0}=\Delta_{\gamma_i\gamma_j}$ and
$\Delta_{\gamma_0}=\Delta_{\gamma_i}$. When plotted against $a^2$,
we expect the splittings to show some curvature for all actions,
because in the scaling regime we expect them to scale with
$\alpha_s^2a^2$ for asqtad and HISQ/tree, and with $\alpha_s a^2$ for
stout. We found that $\Delta_\alpha$ for different actions can be fit
to a polynomial form $f(a^2)=B_2a^2+B_4a^4+B_6a^6$ up to the lattice
spacing $a=0.22$ fm. After performing such fits, we obtained
the splittings $\Delta_\alpha$ as functions of $a^2$ for 
HISQ/tree, asqtad and stout.

Having data for the splittings we set the Goldstone pion mass
to $m_{\gamma_5}=140$ MeV and calculate the root-mean-squared (RMS) 
pion mass as a function of the
lattice spacing:
\begin{equation}
  m^{RMS}_\pi=\sqrt{
  \frac{1}{16}\left(m_{\gamma_5}^2+m_{\gamma_0\gamma_5}^2
  +3m_{\gamma_i\gamma_5}^2+3m_{\gamma_i\gamma_j}^2
  +3m_{\gamma_i\gamma_0}^2+3m_{\gamma_i}^2
  +m_{\gamma_0}^2+m_{1}^2\right)}.
\end{equation}
The results are presented in Fig.~\ref{fig:pion_mass}. 
Curves show the RMS pion mass obtained from fitted values of
$\Delta_\alpha$.
The thickness of each band represents the systematic error 
introduced by varying the end of the fitting interval from
0.17 to 0.22 fm. The symbols correspond to the RMS pion
calculated from the measured pion splittings. For our estimates
below we took the midpoints in each band.

Consider lattice spacing $a=0.15$ fm. On an $N_\tau=8$ lattice it corresponds
to temperature $T=164$ MeV, well in the transition region. At this $a$
the RMS pion mass is 306 MeV for HISQ/tree, 
394 MeV for stout, and 
496 MeV for asqtad. Having the same mass as for HISQ/tree requires
$a=0.102$ fm for asqtad and $a=0.109$ for stout. In other words,
a HISQ/tree simulation  on $N_\tau=8$ at $T=164$ MeV is comparable to an
asqtad simulation on $N_\tau=197.3/164/0.102\simeq11.8$, or a stout
simulation on $N_\tau=197.3/164/0.109\simeq11.0$ at the same $T$.
Thus, we expect HISQ/tree $N_\tau=8$ results to be close to 
stout $N_\tau=10$ and asqtad $N_\tau=12$ results, as far as
the taste symmetry is concerned.
For comparison, if one desires to have 200 MeV RMS pion at $T=164$ MeV,
this requires $a=0.089$ fm for HISQ/tree, $a=0.067$ fm for asqtad, and
$a=0.062$ fm for stout. This translates into the temporal extent
of $N_\tau\simeq13.5$, 18.0, 19.4, respectively, for these actions.
The lattice spacings discussed above are represented with vertical
lines in Fig.~\ref{fig:pion_mass}. These estimates are rather crude,
but are completely in line with the conclusion of~\cite{Bazavov:2010ru}
that a HISQ (or HISQ/tree) 
ensemble with spacing $a$ is comparable to an asqtad
ensemble with $2/3a$.

\begin{figure}
\centering
\includegraphics[width=0.32\textwidth]{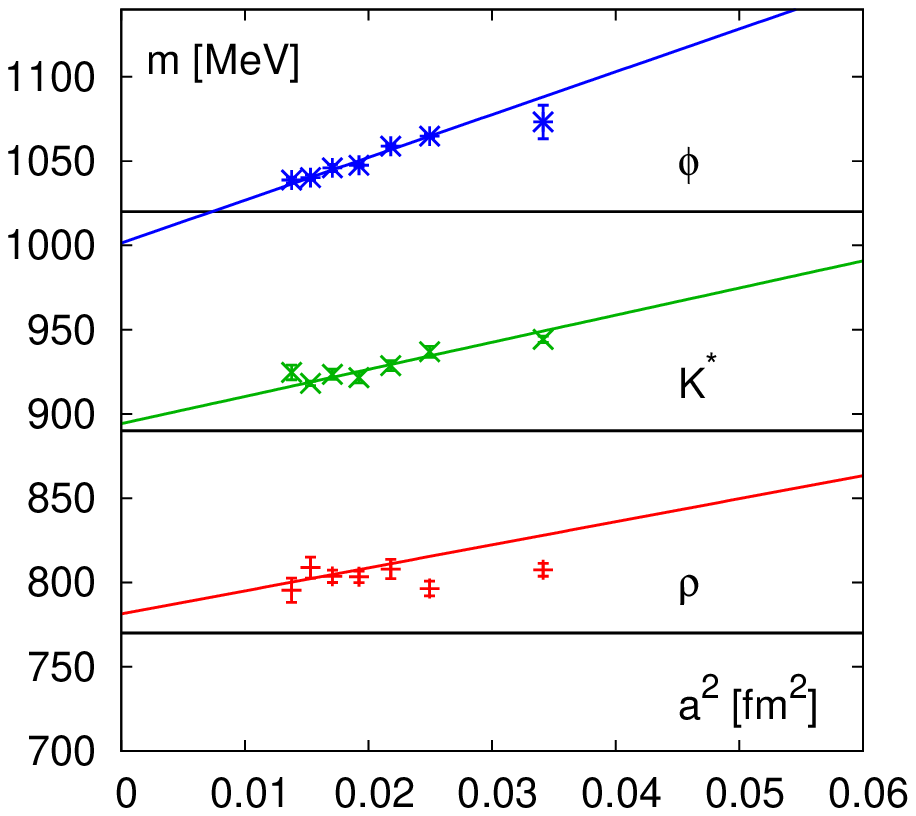}\hfill
\includegraphics[width=0.32\textwidth]{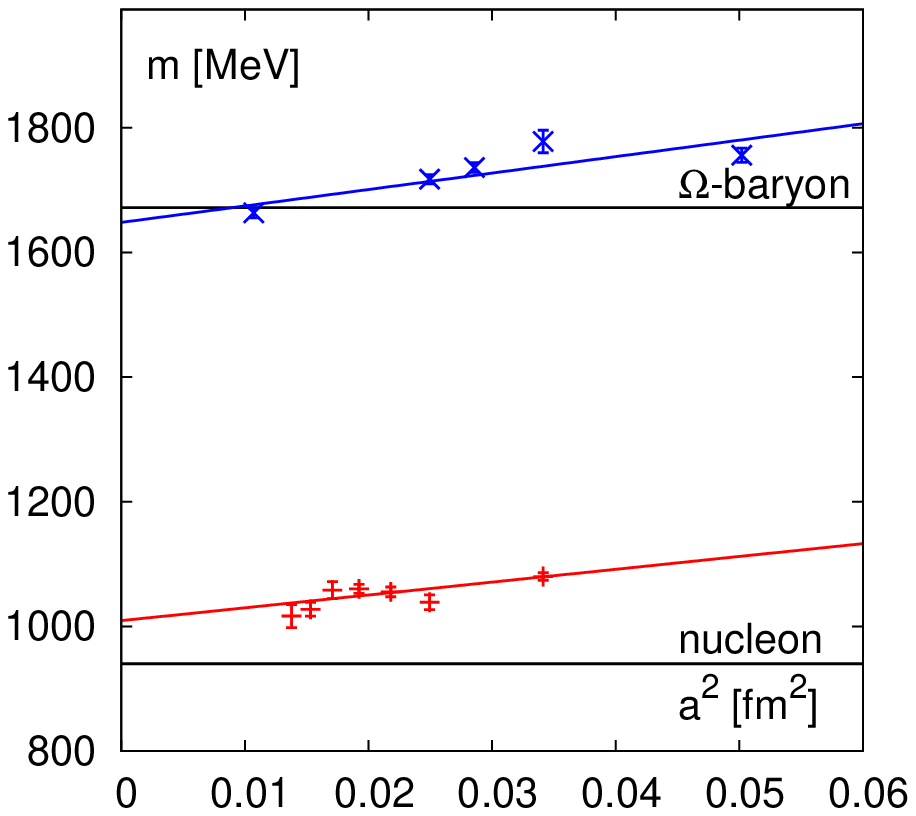}\hfill
\includegraphics[width=0.32\textwidth]{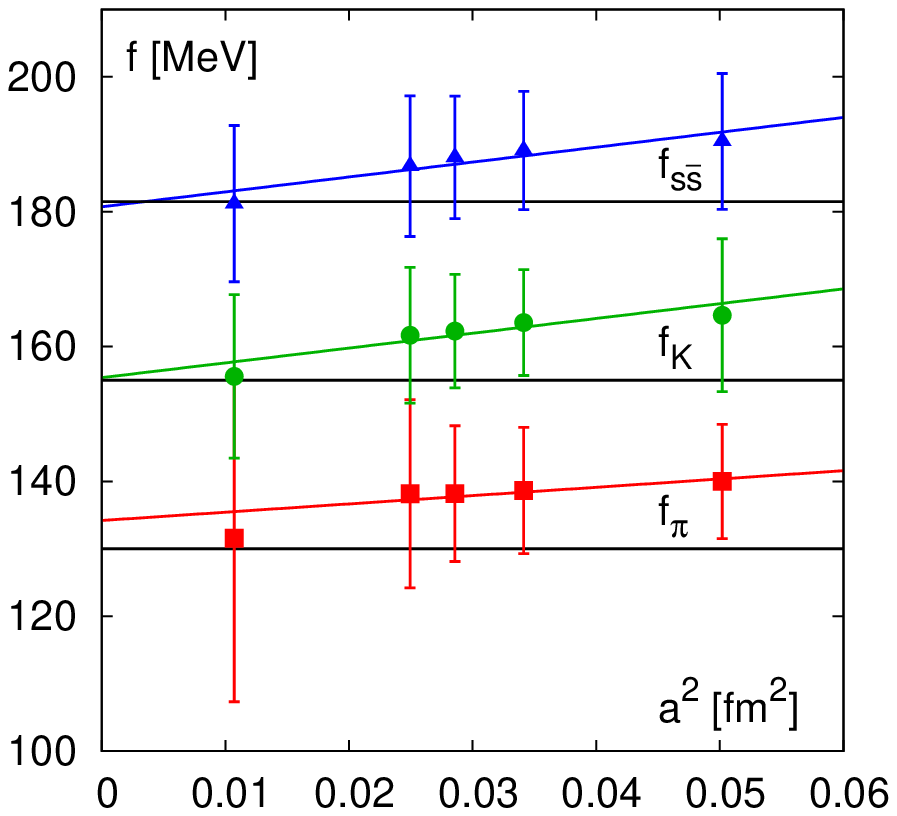}
\caption{Preliminary results for 
masses of vector mesons (left), nucleon and $\Omega$
(middle) and the decay constants of pseudoscalar mesons, including the
fictitious $\eta_{s\bar s}$ meson (right) for HISQ/tree, $m_l=m_s/20$.}
\label{fig:hadrons}
\vspace{-4.0mm}
\end{figure}

Lattice artifacts, related to taste symmetry breaking,
affect masses of hadron states, and,
in general, distort the hadron spectrum. In our simulations the
masses of the pseudoscalar mesons $m_\pi$ and $m_K$ were used as
input to constrain the LCP. However, other states, \textit{e.g.} vector mesons or
baryons, are predictions and can show how well the spectrum 
can be reproduced at a given
lattice spacing. In Fig.~\ref{fig:hadrons} we present masses of
$\rho$, $K^*$, $\phi$ mesons, nucleon and $\Omega$-baryon 
along with $a^2$ extrapolations to
the continuum. We have also calculated the pseudoscalar decay 
constants. Preliminary results for five ensembles are shown
in the rightmost panel of Fig.~\ref{fig:hadrons}.

\section{Taste symmetry and thermodynamic quantities}

\begin{figure}
\centering
\includegraphics[width=0.33\textwidth]{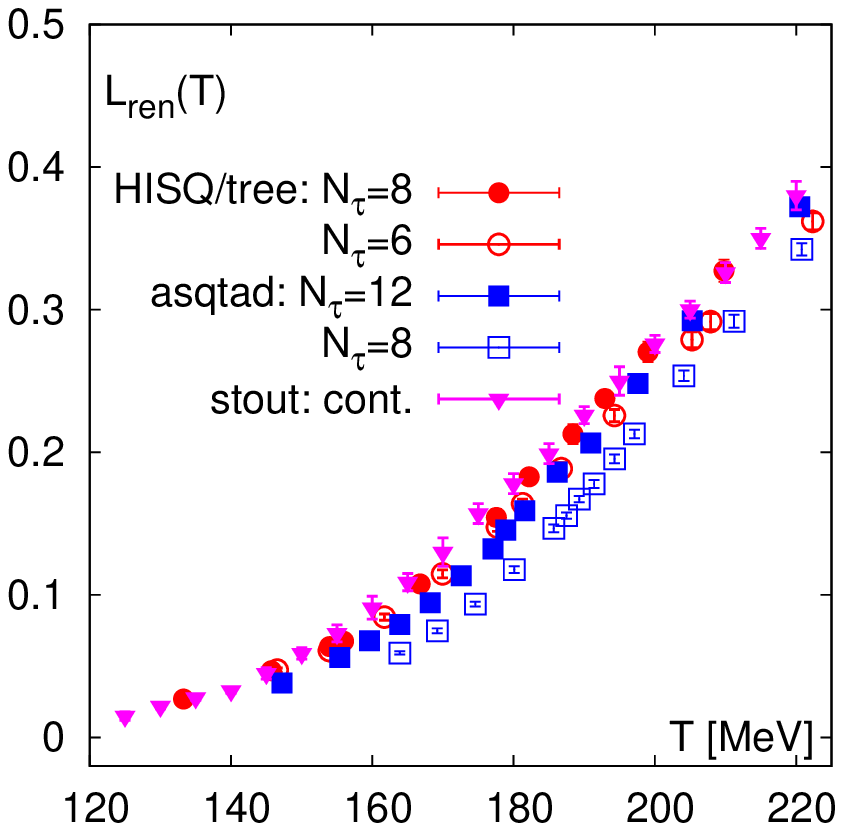}\hfill
\includegraphics[width=0.33\textwidth]{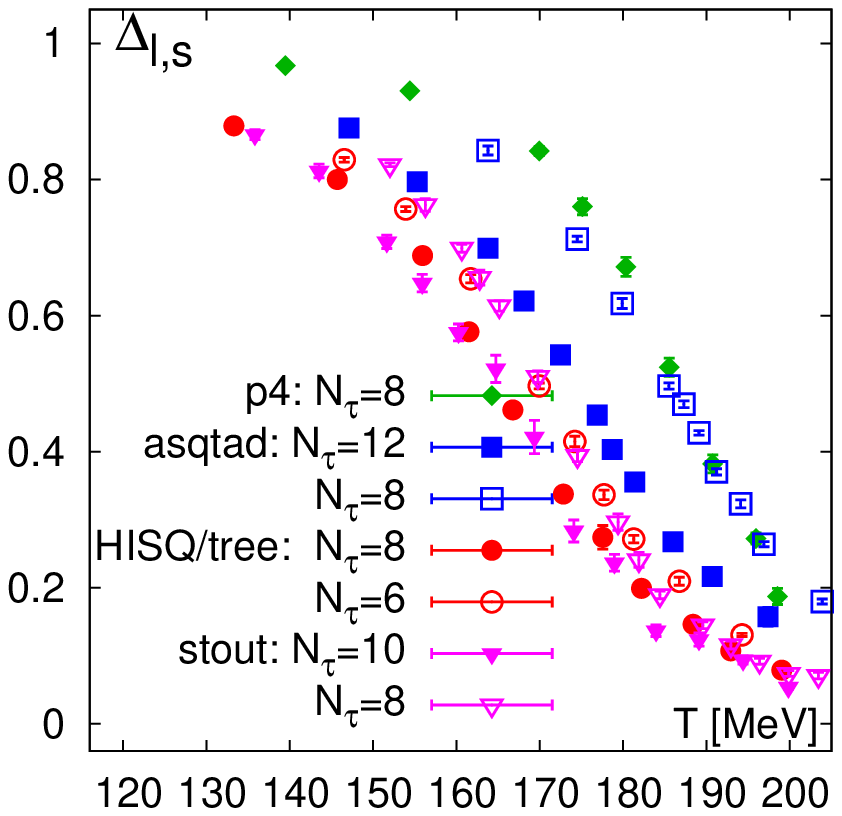}\hfill
\includegraphics[width=0.33\textwidth]{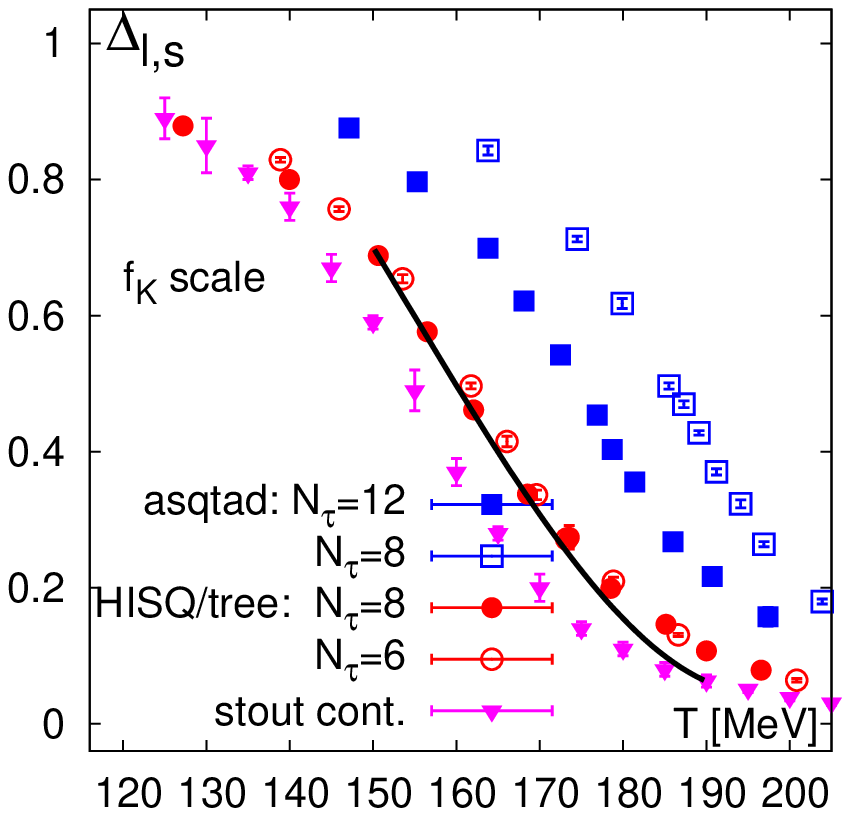}
\vspace{-2mm}
\caption{The Polyakov loop (left), chiral condensate with $r_0$ (middle)
and with $f_K$ (right) scale.}
\label{fig:ploop_and_chiral}
\vspace{-3mm}
\end{figure}

Now we turn to the discussion of different thermodynamic quantities.
In the left panel of Fig.~\ref{fig:ploop_and_chiral} the renormalized
Polyakov loop is shown for several actions. This quantity is 
associated with deconfinement and is not very sensitive to the chiral
properties of the action. Therefore we observe mild
effects from going to finer lattices or to actions having less taste
breaking. The chiral condensate, shown in the middle panel of 
Fig.~\ref{fig:ploop_and_chiral}, on the contrary, exhibits significant
differences for p4, asqtad and HISQ/tree on lattices with the same $N_\tau=8$.
We attribute these to the amount of taste breaking that these actions
have. P4 and asqtad, having higher degree of taste breaking, operate 
at a higher RMS pion mass than HISQ/tree and stout; 
therefore the transition is shifted
to higher temperature. As shown in Sec.~\ref{sec_tastes}
at $T=164$ MeV the asqtad RMS pion
mass is about 496~MeV, and the HISQ/tree mass is about 306~MeV. This improvement
in the taste symmetry shifts the transition region by about 15 MeV.
(Compare HISQ/tree and asqtad $N_\tau=8$ curves in Fig.~\ref{fig:ploop_and_chiral}.)
The stout data in the middle panel of Fig.~\ref{fig:ploop_and_chiral}
is plotted by converting the $f_K$ scale to $r_0$ scale (using $f_K$ and $r_0$
values calculated by the Budapest-Wuppertal collaboration). This allows us
to compare the data for different actions at finite lattice spacing, when
the latter is set by the same procedure.

To directly compare to the results of the Budapest-Wuppertal
group \cite{Borsanyi:2010bp} we have also
plotted the chiral condensate with the temperature scale set by $f_K$
(for HISQ/tree, but not asqtad) rather than $r_0$. This is shown in the right panel of 
Fig.~\ref{fig:ploop_and_chiral}. 
As one can see, HISQ/tree $N_\tau=6$ and 8 results collapse into one curve
which is almost identical with the solid curve that represents the
HISQ/tree continuum result extrapolated from $N_\tau=6$ and 8 using the $r_0$ scale.
We attribute the difference between the HISQ/tree and stout continuum
extrapolations to the light quark mass difference: HISQ/tree (and asqtad)
LCP uses $m_l=m_s/20$, while stout LCP is tuned to $m_l=m_s/27$.

\begin{figure}
\begin{center}
\includegraphics[width=0.48\textwidth]{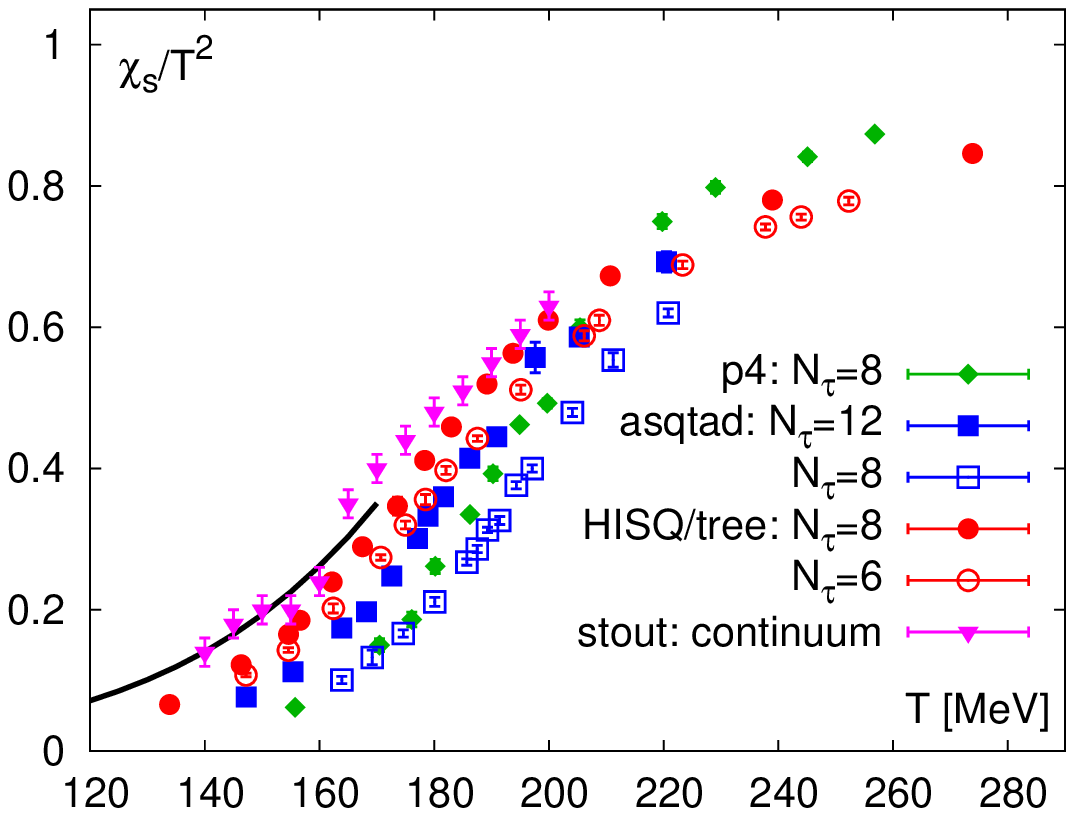}
\hfill
\includegraphics[width=0.48\textwidth]{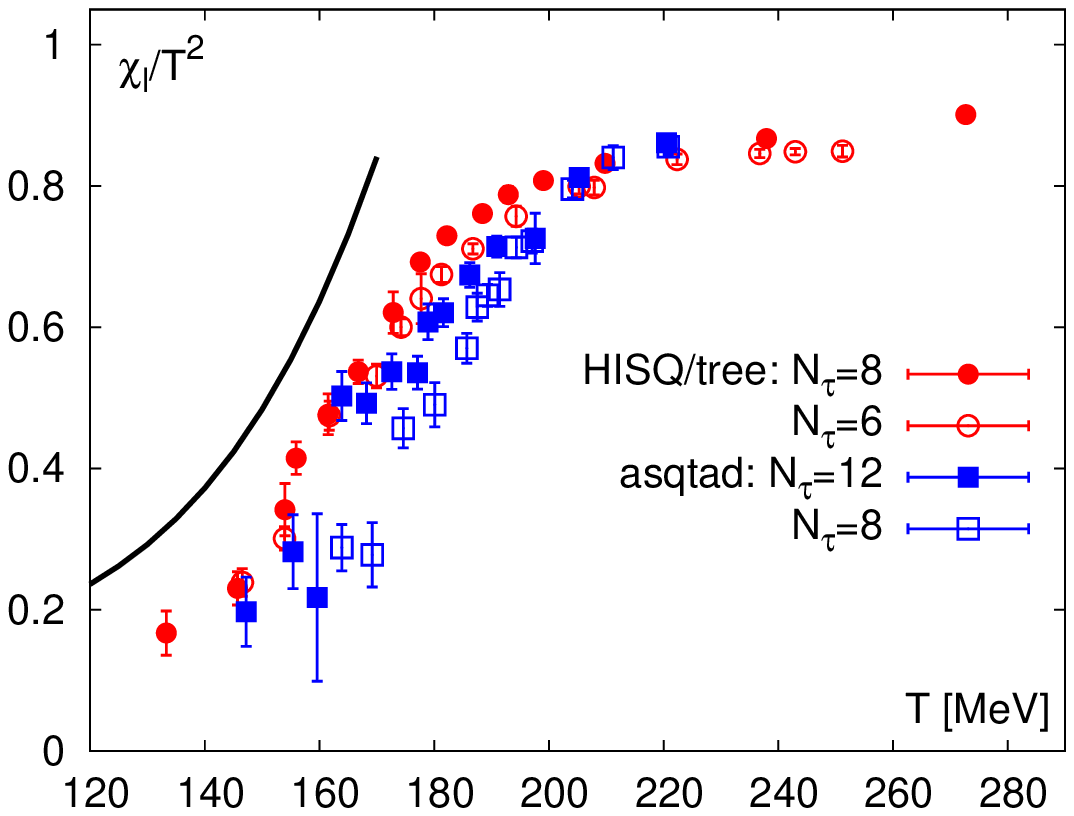}
\vspace{-2mm}
\parbox[b]{0.48\textwidth}{
\caption{The strange quark number susceptibility.
The solid line represents the HRG model.
}
\label{fig:chi_s}
}
\hfill
\parbox[b]{0.48\textwidth}{
\caption{The light quark number susceptibility.
The solid line represents the HRG model.
}
\label{fig:chi_l}
}
\end{center}
\vspace{-4mm}
\end{figure}

\begin{figure}
\begin{center}
\includegraphics[width=0.48\textwidth]{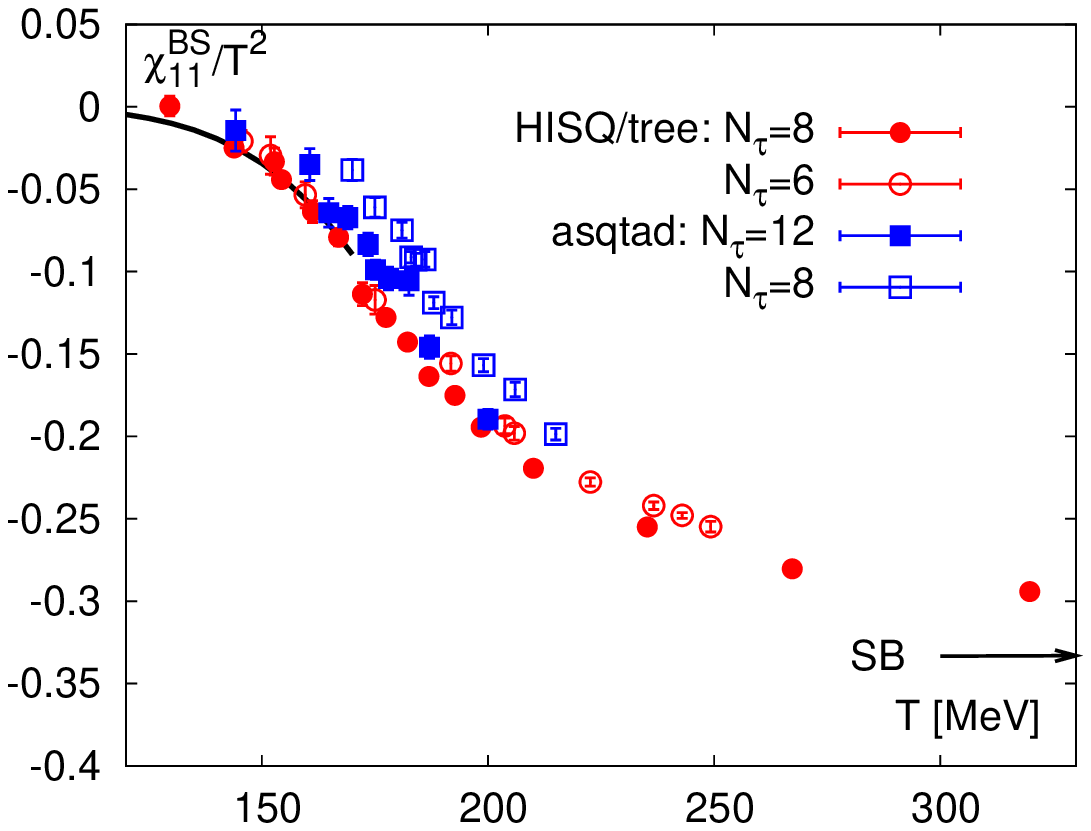}
\hfill
\includegraphics[width=0.48\textwidth]{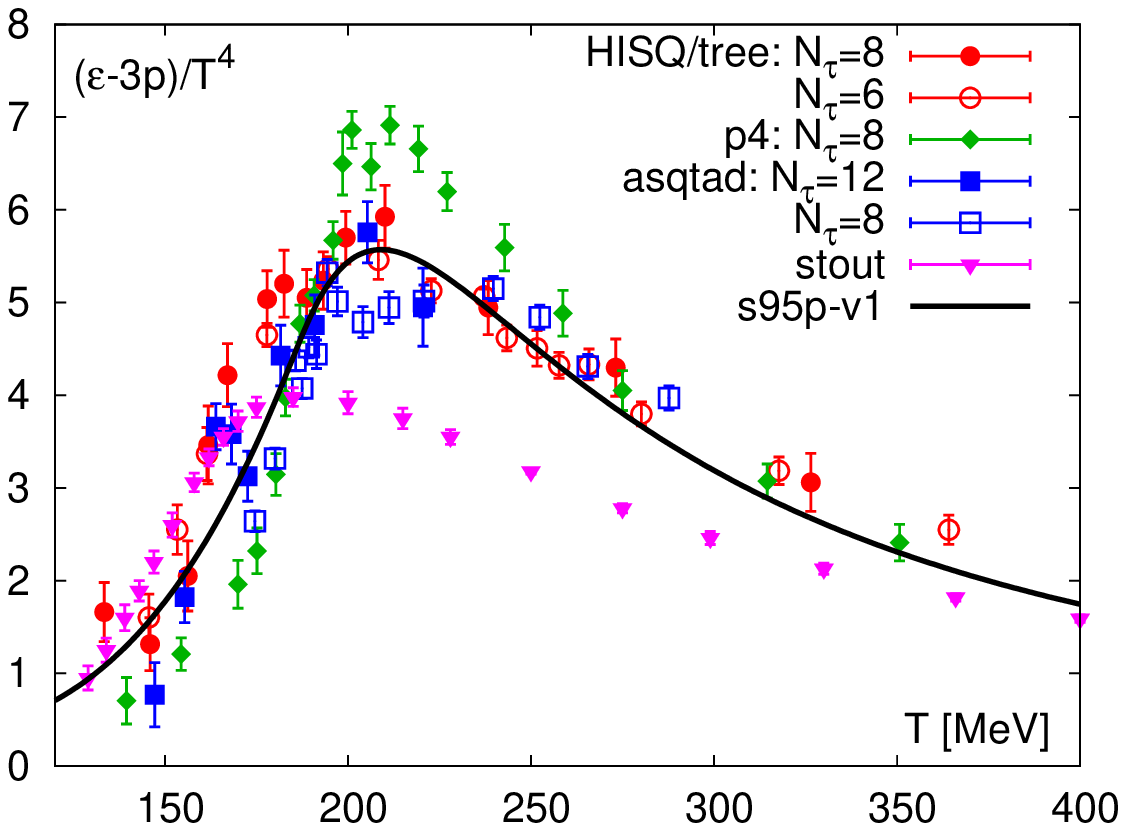}
\parbox[b]{0.48\textwidth}{
\caption{The $S$-$B$ correlations.}
\label{fig:BS_fluc}
}
\hfill
\parbox[b]{0.48\textwidth}{
\caption{The trace anomaly. See details in the text.}
\label{fig:theta}
}
\end{center}
\vspace{-4.5mm}
\end{figure}

We also study the strange and light quark number susceptibilities,
Fig.~\ref{fig:chi_s} and \ref{fig:chi_l}. These quantities are 
associated with deconfinement and signal the liberation of degrees
of freedom with certain quantum numbers. Because these quantities
are related to hadron properties in the low temperature region, 
they are sensitive to the amount of 
taste breaking at finite lattice spacing. In line with previous observations
the asqtad data at finite lattice spacing is shifted to higher 
temperatures, compared to HISQ/tree.
The strangeness-baryon number correlations are shown in
Fig.~\ref{fig:BS_fluc}. The agreement between the HRG model and
the lattice is considerably improved with the HISQ/tree action.
We conclude from Fig.~\ref{fig:BS_fluc} that this quantity
is dominated by the states heavy enough, so that their masses
are reproduced well with the HISQ/tree action even at finite lattice
spacing, corresponding to $N_\tau=6$ and higher.

Finally, we discuss the effects of the taste symmetry breaking
on the trace anomaly, Fig.~\ref{fig:theta}. In the low-temperature
region ($T<160$ MeV) our HISQ/tree results agree well
with the continuum estimate for
the stout data, obtained by Budapest-Wuppertal Collaboration 
by averaging over $N_\tau=6$ and 8 or 
$N_\tau=8$ and 10, respectively 
(see \cite{Borsanyi:2010cj} for details),
but not with the p4 and asqtad results. This is expected,
again, due to higher degree of taste breaking in p4 and asqtad.
At high temperatures ($T>250$ MeV) p4, asqtad and HISQ/tree results
agree between themselves, but significantly disagree with
the stout data.
We note that p4, asqtad, and HISQ/tree actions improve the quark 
dispersion relation whereas the stout action 
(with conventional staggered fermions) does not.
The solid line
in Fig.~\ref{fig:theta} is a parametrization of the equation of 
state derived in \cite{Huovinen:2009yb} based on the hadron resonance
gas model and lattice data.

\section{Conclusion}

Taste symmetry breaking at finite lattice spacing is one of the
major cut-off effects in staggered actions. By comparing
several staggered actions with a different degree of improvement
and at different lattice spacings, we studied the cut-off dependence
of several thermodynamic quantities. As expected, the quantites
that are either directly related to the chiral symmetry
(such as the chiral condensate) or are sensitive to the hadron
spectrum (for instance, quark number susceptibilities and 
the interaction measure) are affected 
by taste symmetry breaking of the staggered fermion actions.
The HISQ action, having the highest degree of improvement,
exhibits the smallest cut-off effects, compared to p4, asqtad and stout.


\begin{thebibliography}{99}

\bibitem{WSlat10}
  W.~S\"{o}ldner [HotQCD Collaboration],
  PoS {\bf LATTICE2010}, 215 (2010)

\bibitem{carleton}
  C.~E.~DeTar,
  PoS {\bf LATTICE2008}, 001 (2008)
  [arXiv:0811.2429 [hep-lat]];
  P.~Petreczky,
  Nucl.\ Phys.\  A {\bf 830}, 11C (2009)
  [arXiv:0908.1917 [hep-ph]];
  P.~Petreczky,
  Nucl.\ Phys.\ Proc.\ Suppl.\  {\bf 140}, 78 (2005)
  [arXiv:hep-lat/0409139].

\bibitem{sharpe06creutz07}
  S.~R.~Sharpe,
  PoS {\bf LATTICE2006}, 022 (2006)
  [arXiv:hep-lat/0610094];
  M.~Creutz,
  PoS {\bf LATTICE2007}, 007 (2007)
  [arXiv:0708.1295 [hep-lat]].

\bibitem{Follana:2006rc}
  E.~Follana {\it et al.}  [HPQCD collaboration and UKQCD collaboration],
  Phys.\ Rev.\  D {\bf 75} (2007) 054502 [arXiv:hep-lat/0610092].

\bibitem{Bazavov:2010ru}
  A.~Bazavov {\it et al.}  [MILC collaboration],
  Phys.\ Rev.\  D {\bf 82}, 074501 (2010)
  [arXiv:1004.0342 [hep-lat]].

\bibitem{hisqthermo}
  A.~Bazavov and P.~Petreczky [HotQCD Collaboration],
  PoS {\bf LATTICE2009}, 163 (2009)
  [arXiv:0912.5421 [hep-lat]];
  A.~Bazavov and P.~Petreczky  [HotQCD collaboration],
  J.\ Phys.\ Conf.\ Ser.\  {\bf 230}, 012014 (2010)
  [arXiv:1005.1131 [hep-lat]].

\bibitem{Borsanyi:2010bp}
  S.~Borsanyi, Z.~Fodor, C.~Hoelbling, S.~D.~Katz, S.~Krieg, C.~Ratti and K.~K.~Szabo
                  [Wuppertal-Budapest Collaboration],
  JHEP {\bf 1009}, 073 (2010)
  [arXiv:1005.3508 [hep-lat]].

\bibitem{Borsanyi:2010cj}
  S.~Borsanyi {\it et al.},
  arXiv:1007.2580 [hep-lat].

\bibitem{Huovinen:2009yb}
  P.~Huovinen and P.~Petreczky,
  Nucl.\ Phys.\  A {\bf 837}, 26 (2010)
  [arXiv:0912.2541 [hep-ph]].


\end{thebibliography}
\end{document}